\documentstyle[epsfig]{europhys}

\def\And{{\rm and\ }}

\def\stars{\bigskip\centerline{***}\medskip}

\newif\ifboo \boofalse

\def\Review#1{\boofalse{\it #1},}
\def\Name#1{{\sc #1},}
\def\Vol#1{\ifboo Vol. {\bf #1}\else{\bf #1}\fi}
\def\Year#1{\ifboo #1\else(#1)\fi}

\def\Page#1{\ifboo {\rm p. #1}\else{\rm #1}\fi}

\begin{document}
\euro{}{}{}{}
\Date{}
\shorttitle{M. GL\"UCK {\it et al.} FRACTAL STABILIZATION OF WANNIER-STARK RESONANCES}

\title{Fractal stabilization of Wannier-Stark resonances}
\author{M. Gl\"uck\inst{1}, A. R. Kolovsky\inst{1,2}
                                          \And H. J. Korsch\inst{1}}
\institute{
     \inst{1} Fachbereich Physik, Universit\"at Kaiserslautern, D-67653 
              Kaiserslautern, Germany\\
     \inst{2} Kirensky Institute of Physics, 660036 Krasnoyarsk, Russia}
\rec{\qquad }{in final form \qquad }

\pacs{
\Pacs{03}{65.$-$w}{Quantum mechanics}
\Pacs{05}{45.$-$a}{Chaotic and nonlinear dynamical systems}
\Pacs{73}{20.Dx}{Electron states in low-dimensional structures}
      }
\maketitle

\begin{abstract}
The quasienergy spectrum of a Bloch electron affected by dc-ac fields 
is known to have a fractal structure as function of the so-called electric 
matching ratio, which is the ratio of the ac field frequency and the Bloch 
frequency. This paper studies a manifestation of the fractal nature of
the spectrum in the system ``atom in a standing laser wave", which
is a quantum optical realization of a Bloch electron.
It is shown that for an appropriate choice of the system parameters
the atomic survival probability (a quantity measured in laboratory
experiments) also develops a fractal structure as a function of the
electric matching ratio. Numerical simulations under classically chaotic
scattering conditions show good agreement with theoretical predictions 
based on random matrix theory.
\end{abstract}
%

\section{1} 
In this letter we study the spectral and dynamical properties of
a Bloch particle affected by static and time-periodic forces:
\begin{equation}
\label{1}
\widehat{H}=\hat{p}^2/2+\cos x+Fx+F_\omega \cos(\omega t)\,x 
\;,\quad \hat{p}=-\hbar'{\rm d}/{\rm d}x\;,
\end{equation}
where $\hbar'$ is the scaled Planck constant (see below).
Originally this problem was formulated for a Bloch electron in dc-ac 
electric fields 
$\widehat{H}=\hat{p}^2/2m+V(x)+e[E+E_\omega\cos(\omega t)]\,x$,
\,$V(x+a)=V(x)$ \cite{niu1} and attracted much attention because of the 
similarity with the Hofstadter problem \cite{hofstadter}. Indeed, the energy
spectrum of a Bloch electron in a 2D lattice under the action of a
constant magnetic field $B$ depends on the magnetic matching ratio
$\beta=h/eBa^2$ ($a$ is the lattice constant) and has a fractal structure
as function of this parameter. Analogously, the quasienergy spectrum
of system (\ref{1}) depends on the electric matching ratio
\begin{equation}
\label{2}
\alpha=\frac{\omega}{\omega_B} \;,\quad \omega_B=\frac{2\pi F}{\hbar'} \;,
\end{equation}
where $\omega_B$ is the Bloch frequency, which
is $\omega_B=eEa/\hbar$ in the case of a crystal electron.
For rational ratios $\alpha=r/q$ the quasienergy spectrum has a band
structure; it is discrete, however, for irrational values of $\alpha$
\cite{niu1}. Numerically the quasienergy
spectrum of the system (\ref{1}) was studied in ref.~\cite{niu2} by 
using the tight-binding approximation. It was show that
for $\alpha=r/q$ the quasienergy bands are arranged in a structure 
resembling the famous Hofstadter butterfly (see Fig.~1 in Ref.~\cite{niu2}).
It should be pointed out, however, that the results of paper \cite{niu2} 
only partially describe the spectrum of the system (\ref{1}) 
because the tight-binding approximation
neglects the decay of the quasienergy states. The actual quasienergy
spectrum is complex, where the imaginary part of the spectrum defines the 
lifetimes of the metastable quasienergy states.

Because of the extremely small lattice period in crystals, the fractal 
structure of the quasienergy spectrum has never been observed in solid state
systems. However, a signature of it was recently found in an experiment
with cold atoms in an optical lattice \cite{raizen2}. The latter system
models the solid state Hamiltonian (\ref{1}), where the neutral atoms
moving in the optical potential $V(x)$ ($k_L$ is the laser wave vector) 
take over the role of the crystal electrons. The effect of
the electric fields can be mimicked, for example, by the inertial force 
induced by accelerating the experimental setup as a whole. 
(In practice, however, the acceleration  was obtained by an
appropriate chirping of the laser frequency.)
The system ``atom in a standing wave" has an essentially larger lattice
period than the solid state system and is, in addition, free
of relaxation processes due to scattering by impurities and the Coulomb
interaction. These features of the system were utilized earlier in 
ref.~\cite{raizen1} to observe experimentally the Wannier-Stark ladder 
of resonances. The main modification of the experiment \cite{raizen2} in
comparison with the experiment  \cite{raizen1} is that a strong
periodic driving with frequency $\omega=(r/q)\omega_B$
was imposed. (Only the cases $r/q=1/2$ and $r/q=1/3$ were reported.)
Then the atomic survival probability as a function of the frequency
of the probe signal shows additional anti-peaks (see fig.~3 in 
ref.~\cite{raizen2}), which were interpreted as an indication of 
the fractal structure of the spectrum. We note that the width of these 
anti-peaks is given by the width ({\it i.e.}~the inverse lifetime) 
of the first excited Wannier-Stark resonances.
This imposes a fundamental restriction on the resolution and it seems
impossible to resolve matching ratios $\alpha=\omega/\omega_B$
for $\alpha$ different from lowest rational numbers.


\section{2}
The discussed papers \cite{niu2,raizen2} study the  system
in the deep quantum region. In the present paper we discuss
the manifestation of the fractal structure of the quasienergy spectrum
in the semiclassical region of the system parameters. 
We shall show that in this region the fractal nature of the spectrum 
can be observed without using a probe signal and the electric matching
ratio can be resolved with any desired accuracy.

The characteristic measure of the systems ``classicality" is the scaled Planck
constant $\hbar'$ entering the momentum operator in the Hamiltonian (\ref{1}).
Referring to the system ``atom in a standing wave" 
the scaled Planck constant is given by
\begin{equation}
\label{3}
\hbar'=\left(\frac{8\hbar\omega_{rec}}{V_0}\right)^{1/2} \;,
\end{equation}
where $\omega_{rec}=\hbar k^2/2m$ is the atomic recoil frequency and 
$V_0$ is the depth of the optical potential $V(x)=V_0\cos^2(k_L x)$. 
In the experiments \cite{raizen2} and \cite{raizen1} the value of 
the scaled Planck constant was $\hbar'\approx1.5$ and $1.6$, respectively. 
In our numerical studies we use $\hbar'=0.25$. Since the amplitude $V_0$ 
of the optical potential is proportional to the square of the laser field 
amplitude, this implies a larger intensity of the laser.
 
We simulate the wave packet dynamics of the system (\ref{1})
by numerical solution of the time-dependent Schr\"odinger equation
in the momentum representation using parameters
$\hbar'=0.25$, $\omega=10/6$, $F_\omega=4.16$ with values of
the static field $F$ in the interval $0.030\le F\le0.083$.
Then the survival probability
\begin{equation}
\label{3a}
P(t)=\int_{|p|<p_0} |\psi(p,t)|^2 {\rm d}p \;,
\end{equation}
is calculated. The initial state is the localized Wannier state, which for 
the cosine potential practically coincides with a minimal uncertainty wave 
packet centered at $x=\pi$. The motivation for the chosen value of
$p_0=6$, which is much larger than the region of support of the initial
wave packet, is given in sec.~3 below. This numerical simulation models the 
experimental situation of ref.~\cite{raizen2} with the modification that 
there is no probe signal but the amplitude $F$ of the static force is varied.

\begin{figure}
\begin{center}
\includegraphics[height=8.5cm,clip]{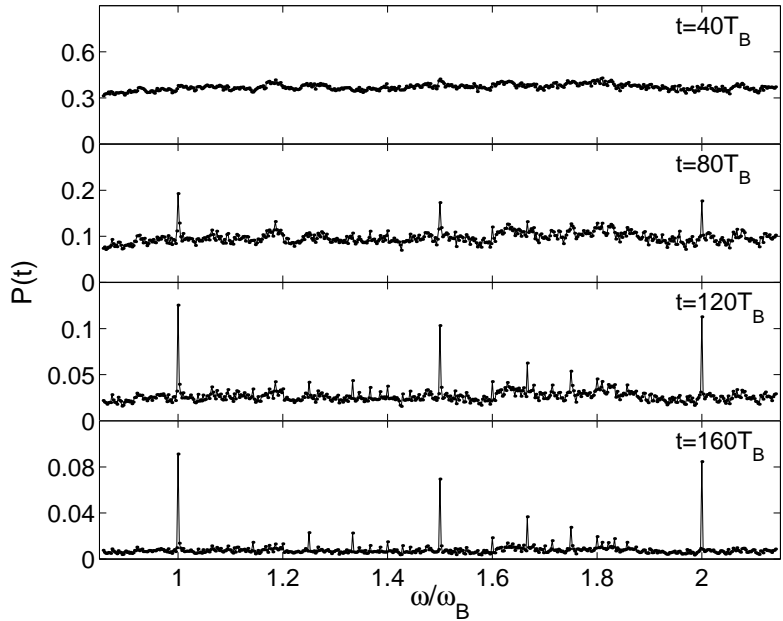}
\caption{Time evolution of the atomic survival probability $P(t)$ calculated
for rational values of the electric matching ratio $\omega/\omega_B=r/q$, 
$q\le 420$ (see text). With increasing time 
$t=40\,T_B$,\ldots $t=160\,T_B$ the survival probability develops
pronounced peaks at rationals with small denominators.  
The system parameters are $\hbar'=0.25$, $\omega=10/6$,
$F_\omega/\omega^2=3/2$, and $0.031<F<0.077$.}
\label{fig1}
\end{center}
\end{figure}

The result for the survival probability as a function of $F$ --- or,
equivalently, as a function of the matching ratio $\alpha =
\omega /\omega_B$ --- is shown 
in fig.~\ref{fig1} and \ref{fig2}(a), which is the central result of the paper.
The dots in fig.~\ref{fig1} mark the survival probability $P(t)$ calculated
for increasing values of the observation time $t=40\,T_B$,\ldots $t=160\,T_B$.
The values of static force $F$ are chosen 
for rational values of the electric matching ratios $\omega/\omega_B=r/q$
with $q\le 420$. (Explicitly, we considered for $q$ all divisors of 
$2^2\cdot 3\cdot 5\cdot 7=420$ and all coprime $r$-values
in the interval $6/7\le r/q \le 15/7$.)
To guide the eyes, the dots are connected 
by a solid line. The actual width of the peaks, which is inversely 
proportional to the observation time, is smaller in
fig.~\ref{fig2}(a), which shows a more fully developed distribution
for a longer time $t=200\,T_B$.
It is seen in fig.~\ref{fig2}(a) that there are pronounced 
peaks above an almost constant background which
appear at low order resonances between the driving frequency and
the Bloch frequency. The seven largest peaks in the figure
correspond (from left to right) to $r/q=1$, $5/4$, $4/3$, 
$3/2$, $5/3$, $7/4$, and $r/q=2$, and the observed peak-heights
fall off rapidly with increasing denominator $q$.  
As illustrated by fig.~\ref{fig1},
this peak structure develops gradually in time, starting
from $P(0)=1$ at time zero and originates from the different
long-time behavior of the survival probabilities. We find
exponential decay in time for irrational values of $\omega/\omega_B$
and algebraic decay for rational ones \cite{PE}. Thus the survival 
probability as a function of $F$ reflects the fractal nature of the
spectrum in the long-time regime. Moreover, there
is no fundamental resolution constraint and an arbitrary number of peaks
can be resolved by increasing the observation time. We also note that
instead of varying the static force one can vary the frequency $\omega$ 
of the driving force. In this case, having in mind a laboratory experiment,
it looks reasonable to use the gravitational force as the static force 
\cite{kasevich}.


\section{3} 
The key point of the conducted numerical experiment is that 
the amplitude $F_\omega$ and the frequency $\omega$ of the driving
force are chosen such as to insure chaotic dynamics 
of the system in the classical limit. We furthermore choose 
$p_0$ in (\ref{3a}) to be larger than the boundary between the 
regular and the chaotic component of the classical phase space 
(see fig.~1(b) in ref.~\cite{PLA1}). Then the classical survival 
probability decays exponentially \cite{PLA1} 
\begin{equation}
\label{4}
P_{cl}(t)=\exp(-\nu t) \;,
\end{equation}
where the decay coefficient $\nu$, the inverse classical lifetime,
is determined by the classical Lyapunov exponent and the fractal
dimension of the chaotic repellor. We computed the classical decay rate
numerically and determined the $F$-dependence of the classical
decay rate as $\nu\approx 0.15\,F$ in the parameter region considered here.
It should be noted that the exponential decay of the
classical probability cannot be considered as an universal law.
In some systems it is a transient phenomenon and changes to an algebraic 
decay caused by long-lived trajectories sticking near stability
islands \cite{klafter,chirikov}. However, this is not the case for system
(\ref{1}) and no sign of an algebraic decay of the classical probability
was detected (at least until time $t=200T_b$, which was the maximal
time in our numerical simulation).

\begin{figure}
\begin{center}
\includegraphics[width=7cm,height=6cm,clip]{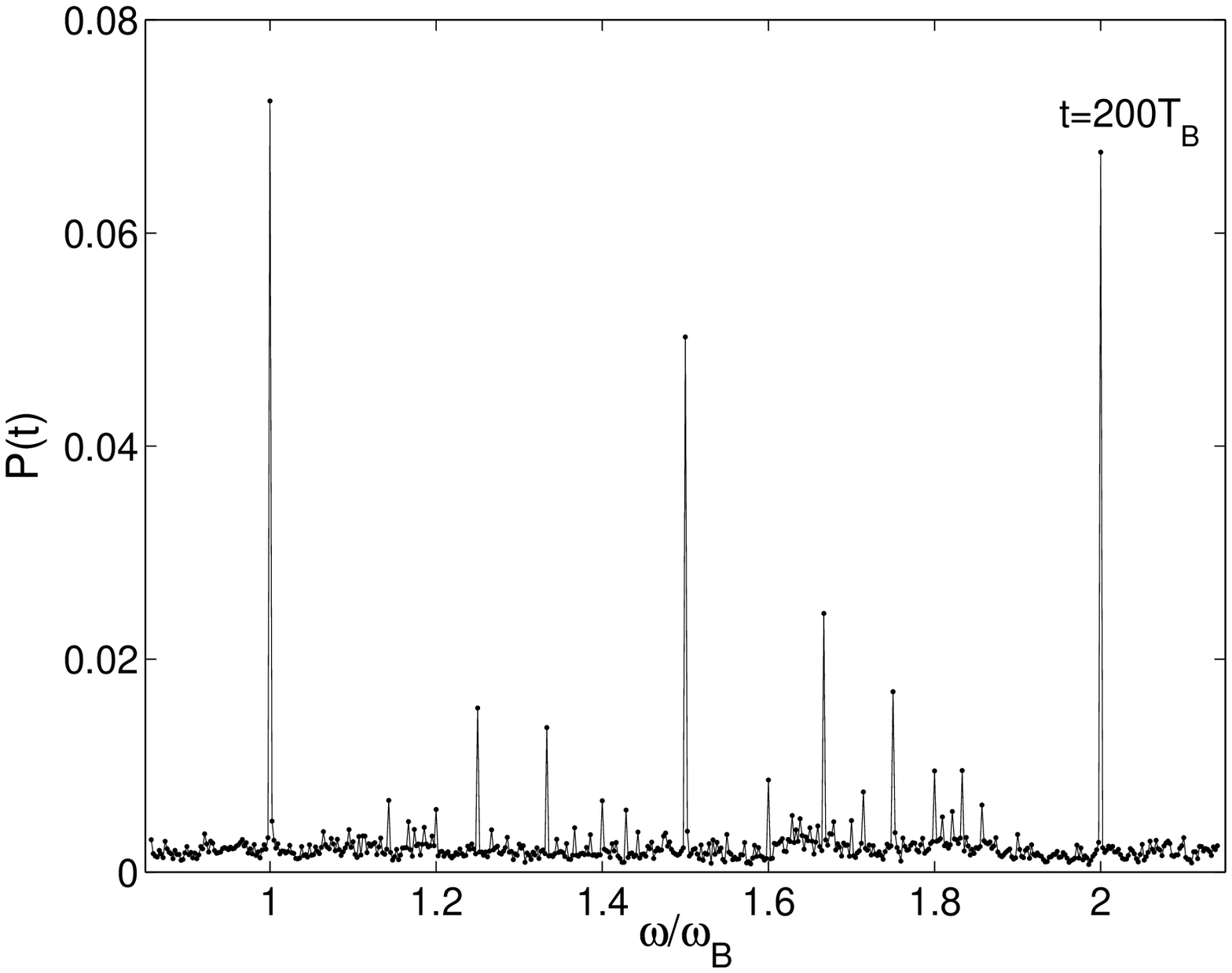}
\includegraphics[width=7cm,height=6cm,clip]{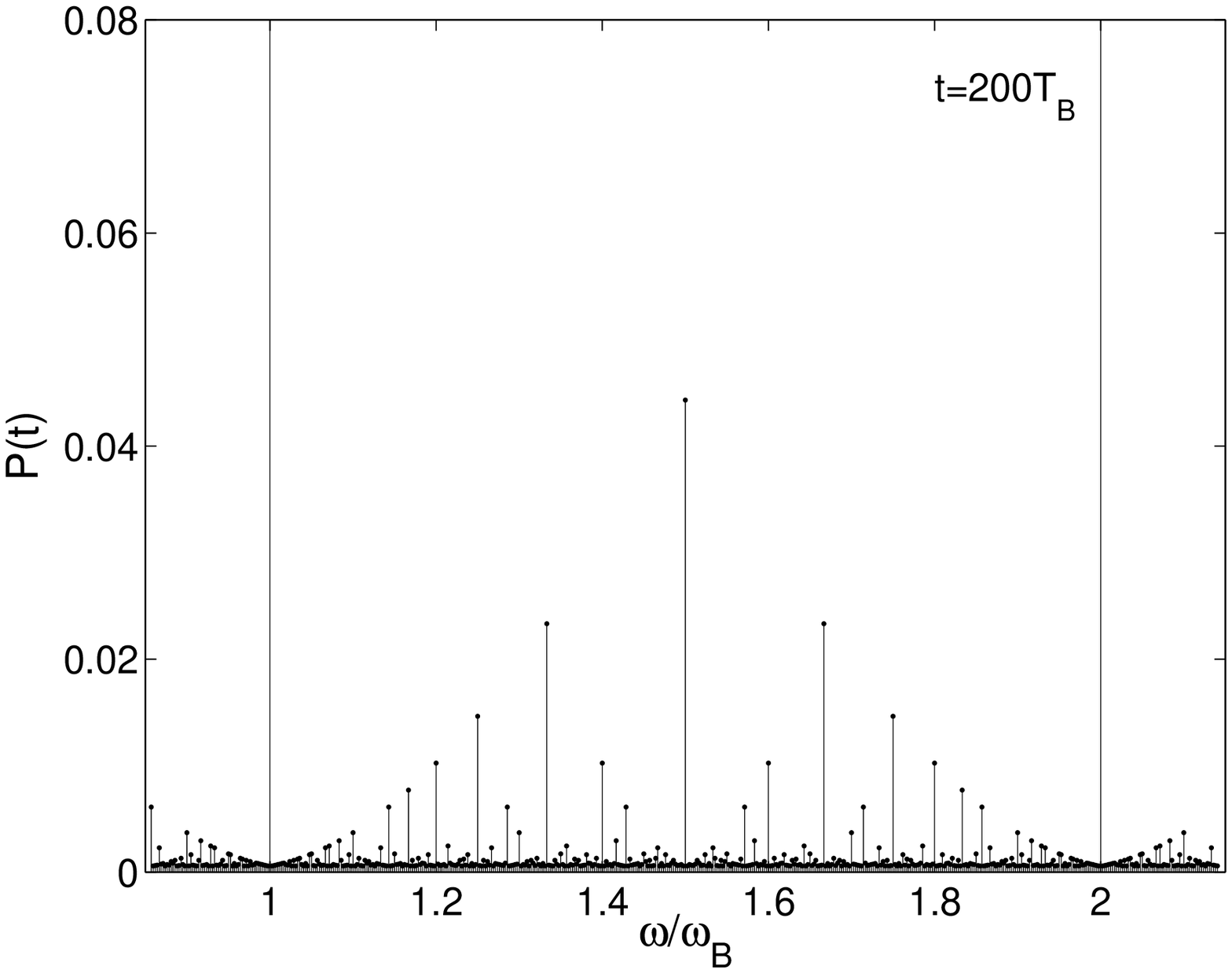}
\caption{(a) Atomic survival probability $P(t)$ as shown in fig.~\ref{fig1},
however for a longer time $t=200\,T_B$, where the fractional structure
is much more clearly developed. To guide the eyes the dots are connected 
by a solid line but the actual width of the peaks, which is inversely 
proportional to the observation time, is smaller.
(b) Survival probability $P(t)=(1+\nu t/q)^{-q}$ (eq.~(\ref{6}))
derived from random matrix theory for $t=200\,T_B$. 
The value of the classical decay rate is $\nu=0.15\,F$.}
\label{fig2} 
\end{center}
\end{figure}

Because we measure the observation time $t$ in units of the Bloch
period $T_B=\hbar'/F$, the $F$ dependence in Eq.~(\ref{4}) cancels and the
classical survival probability is practically constant.
The quantum results follow closely the classical exponential decay 
(\ref{4}) for irrational values of the matching ration $\omega/\omega_B$ 
providing the flat background of the quantum results shown in fig.~\ref{fig2}. 
The peaks above the classical plateau for resonant driving, 
{\it i.e.}~rational values of the matching ratio $\alpha=\omega/\omega_B$ 
are a quantum ``stabilization"-phenomenon, which can be understood 
as follows.

It was shown in ref.~\cite{PRE2} that the eigenvalue problem
for the quasienergies of the system (\ref{1}) for $\alpha=r/q$ can be
mapped onto an effective scattering problem with $q$ open channels.
When the matching ratio $\alpha$ is an irrational number, the number of
channel is infinite and the system follows the classical dynamics
provided the condition $\hbar'\ll 1$ is satisfied. When the matching ratio 
is a rational number, however, the number of decay channels is 
finite and the quantum system (independent of the value of $\hbar'$) 
is essentially more stable than the classical one \cite{PLA1}.
In this case the behavior of the survival probability $P(t)$ differs from 
the exponential decay (\ref{4}) and is determined by the
distribution of the imaginary parts of the quasienergies, 
{\it i.e.}~the distribution of the resonance widths.

In the case of chaotic classical dynamics an analytic expression 
for the resonance statistics is supplied by (non-hermitian) random
matrix theory (RMT) \cite{feydorov,zyczkowski}. The validity of RMT for system 
(\ref{1}) was checked numerically in ref.~\cite{PRE2,PE} and a satisfactory 
correspondence was noticed. Converting the result of RMT from energy 
to time domain shows that the decay of the probability follows asymptotically 
an inverse power law \cite{sokolov}
\begin{equation}
\label{5}
P(t)\approx (\Gamma_W t/q\hbar')^q \;,\quad t\gg t^*\approx q\hbar'/\Gamma_W \;,
\end{equation}
where $\Gamma_W$ is the Weisskopf width, which is a free parameter in the
abstract RMT. Identifying the parameter $\Gamma_W/\hbar'$ with the classical
decay coefficient $\nu$, eq.~(\ref{4}) and eq.~(\ref{5}) can be combined 
in the single equation
\begin{equation}
\label{6}
P(t)=\left(1+\frac{\nu t}{q}\right)^{-q} \;,
\end{equation}
which has the correct short- and long-time asymptotic and
provides a first crude approximation to the more elaborate
RMT result \cite{sokolov}.

Figure \ref{fig2}(b) shows the values of the function (\ref{6}) 
for $t=200\,T_B$ and the same values of the matching ratio $\alpha$ 
as in fig.~\ref{fig2}(a) where we use a slightly different 
graphic presentation of $P(t)$ to stress that the
function (\ref{6}) is a discontinuous function of $\alpha$ for
any $t$. In contrast, the atomic survival probability  shown in
fig.~\ref{fig2}(a) is a continuous function of $F$ where its fractal structure
develops gradually as $t\rightarrow\infty$. 
In fact, the probabilities (\ref{3a}) calculated for two close
rational numbers $\alpha_1$ and $\alpha_2$ follow each other
during a finite ``correspondence" time. (For instance, for
$\alpha_1=1$ and $\alpha_2=999/1000$ the correspondence time
is found to be about $50\,T_B$.) Thus it takes some time to distinguish two
close rationals, although they may have very different 
denominators and, therefore, very different asymptotics (\ref{5}).
With this remark reserved, a nice structural (and even semiquantitative) 
correspondence is noticed. 
In addition, it seems worthwhile to note that also the pronounced
quantum resonance peaks, {\it i.e.}~the quantum algebraic decay 
$P(t)\approx (\nu t/q)^{-q}$ predicted by RMT, is mainly determined by
the purely classical $\nu $-coefficient due to classically chaotic 
scattering dynamics. 


\section{4} 
We have analyzed the system (\ref{1}) in context with recent experiments 
studying the dynamics of cold atoms in a standing laser wave \cite{raizen2}.
It is shown that in the semiclassical region of the system parameters
the atomic survival probability as a function of the static
force (or, alternatively, of the driving frequency) shows a fractal 
structure. This fractal structure is actually related to the
fractal nature of the quasienergy spectrum determined by the
degree of rationality of the electric matching ratio (\ref{2}). 
In fact, when a rational sequence of $\alpha=r/q$ converges to some 
irrational value, the quasienergy bands progressively split into sub-bands. 
This process is accompanied by loss of stability of the quasienergy states 
and shows up, finally, in the complicated (fractal) structure of the atomic 
survival probabilities which can be measured in laboratory experiments.

Finally, we would like to distinguish the fractal structure of
the survival probability discussed above from the fractal structure of the
survival probability studied in papers \cite{ketzmerick,casati,lewenkopf}. 
The latter appears as a quantum manifestation of the hierarchical island
structure of classical phase space in a system with an algebraic decay of the 
classical probability. The origin of the former phenomenon, however, is the
fluctuating number of the decay channels depending on the value of
the electric matching ratio (\ref{2}), which is the control parameter
of the system (\ref{1}).

\stars
This work has been supported by the Deutsche Forschungsgemeinschaft
(SPP 470 {\it `Zeit\-ab\-h\"an\-gige Ph\"anomene und Methoden in 
Quantensystemen der Physik und Chemie'\/}). 

%

%
\end{document}